\documentclass[useAMS,usenatbib]{mn2e}

\title[Monthly Notices: On GRCs problem within an axisymmetric approach]
  {On the galactic rotation curves problem within an axisymmetric approach}

  \author[A. Herrera-Aguilar et al.]
  {A.~Herrera-Aguilar,$^1$$^,$$^2$\thanks{E-mail: aha@fis.unam.mx}
  U.~Nucamendi,$^2$\thanks{E-mail: ulises@ifm.umich.mx} E.~Santos,$^3$\thanks{E-mail: eli@unach.mx} O.~Corradini,$^4$\thanks{E-mail: olindo.corradini@gmail.com}
  \newauthor 
  and C.~Alvarez$^4$\thanks{E-mail: cesar.alvarez@unach.mx} \\
  $^1$Instituto de Ciencias F\'{\i}sicas, Universidad Nacional Aut\'onoma de M\'exico\\
  $^2$ Instituto de F\'{\i}sica y Matem\'{a}ticas,Universidad Michoacana de San Nicol\'as de Hidalgo\\
  $^3$International Centre for Theoretical Physics, Meso American Institute for Science, Universidad Aut\'onoma de Chiapas  \\
  $^4$CEFyMAP, Universidad Aut\'onoma de Chiapas}

  \date{Released 2002 Xxxxx XX}

\pagerange{\pageref{firstpage}--\pageref{lastpage}} \pubyear{2002}

\def\LaTeX{L\kern-.36em\raise.3ex\hbox{a}\kern-.15em
    T\kern-.1667em\lower.7ex\hbox{E}\kern-.125emX}

\begin{document}

\label{firstpage}

\maketitle

\begin{abstract}
In  (\citealt{nucamendi}; \citealt{lake}) it has been shown that galactic potentials can be
kinematically linked to the observed red/blue shifts of the corresponding galactic rotation curves
under a minimal set of assumptions: the emitted photons come from stable timelike circular
geodesic orbits of stars in a static spherically symmetric gravitational field, and propagate to
us along null geodesics. It is remarkable that this relation can be established without appealing
at all to a concrete theory of gravitational interaction. Here we generalize this kinematical
spherically symmetric approach to the galactic rotation curves problem to the stationary
axisymmetric realm since this is precisely the symmetry that spiral galaxies possess. Thus, by
making use of the most general stationary axisymmetric metric, we also consider stable circular
orbits of stars that emit signals which travel to a distant observer along null geodesics and
express the galactic red/blue shifts in terms of three arbitrary metric functions, clarifying the
contribution of the rotation as well as the dragging of the gravitational field. This stationary
axisymmetric approach distinguishes between red and blue shifts emitted by circularly orbiting
receding and approaching stars, respectively, even when they are considered with respect to the
center of a spiral galaxy, indicating the need of precise measurements in order to confront
predictions with observations. We also point out the difficulties one encounters in the attempt of
determining the metric functions from observations and list some potential strategies to overcome
them.
\end{abstract}


\section{Introduction}

The galactic rotation curves provide a direct method of determining the gravitational field inside
a spiral galaxy since they have been measured for a great amount of galaxies (\citealt{rubin1};
\citealt{rubin1a}; \citealt{rubin2}; \citealt{rubin1b}; \citealt{persic}; \citealt{sofue}). These
curves are obtained by measuring the red/blue shifts of light emitted from stars and from the 21
cm radiation from neutral gas clouds. The observations show evidence that the red/blue shifts $z$,
or equivalently, the tangential velocities of rotation v, remain constant or decay more slowly
than the Keplerian behaviour ($v^2\sim 1/r$) up to distances far beyond the luminous radius of
these galaxies. By performing a naive Newtonian analysis of this effect, one deduces that the
energy density of the galaxies decreases approximately as $r^{-2}$, and hence, the mass of these
bodies should increase as $m(r) \approx r$. Since the observed luminous galactic components do not
produce this growing behaviour, a question arises: what is the reason of such an effect? Nowadays
there is a strong believe that dark matter is responsible for it, being the major bounded
constituent of galaxies and galaxy clusters ($\sim 25 \%$ of the total energy density of our
Universe \citep{komatsu}). However, one can naturally ask whether this large unseen mass does not
produce a relevant gravitational red shift.  On the other hand, in (\citealt{salucci}) it was
shown for a large and complete sample of spiral galaxies that their luminous regions consist of
stellar discs embedded in universal dark halos of constant density independently of the galaxy
properties; moreover, it was shown that the Dark Matter halos made by the most likely dark
particle are inconsistent with actual observations. Finally, alternative approaches to this
problem like modifications of Newtonian dynamics based on these observations have also been
developed (\citealt{mond}; \citealt{milgromsanders}; \citealt{lobo}; \citealt{mendoza};
\citealt{hernandez}). Here, following \citep{nucamendi}, we shall analyze the problem on the basis
of what it is directly observed: the red/blue shifts. This approach enables us to keep track of
the effect of the underlying made assumptions, and to be aware of when they are not longer valid.

There have been previous approaches to the galactic rotation curves problem that make use of
stationary axisymmetric metrics within the framework of General Relativity (see the comoving dust
solution of (\citealt{CT}; \citealt{CTb}; \citealt{CTc}; \citealt{CTd}; \citealt{CTf};
\citealt{CC})). Within these models it is claimed that the known data from galactic rotation curve
can be described with at most relatively little extra matter by non-linear general relativistic
effects in galactic dynamics. An important point within their approach is that even though fields
and velocities are small in a galaxy, it is not consistent to describe the latter through a
Newtonian approximation since there are non-linear contributions of non-negligible size coming
from Einstein equations. However, these models have received criticisms in several directions that
point out to unphysical features like the need of additional exotic matter source in the galactic
disc (\citealt{korzynski}) or infinite mass at large distances (\citealt{menzies}), the presence
of singularities when continuing the interior solution into a consistent exterior configuration
(\citealt{zingg}), among others, as well as the inconsistency of the use of comoving frames with
the condition of differential rotation (\citealt{cross}). An important observation upon this model
was pointed out in \citep{RS}, namely, the stationary axisymmetric metric of the comoving dust
solution of (\citealt{CT}) does not possess the most general form since it has only three
arbitrary functions instead of four, moreover, their assumed Weyl gauge (W=r) is not consistent
with the Einstein equations since this metric function is not harmonic and hence, it does not
belong to the most general class of stationary axisymmetric solutions, the Lewis-Papapetrou class,
a circumstance that might be connected with the problems of the model.

The aim of this paper is to provide a stationary axisymmetric kinematical description of the
galactic rotation curves problem since this is a more realistic symmetry compared to the spherical
one when studying spiral galaxies: the most accepted composition of spiral galaxies indicates that
its main mass constituent is concentrated in a thin disc with a central bulge which are surrounded
by a spherical halo (\citealt{bin}). Thus, it is commonly accepted that the main aspects of the
galactic dynamics can be approximately described by (rotating) thin galactic disc models. We make
use of the most general stationary axisymmetric metric and express the galactic red/blue shifts
measured by a distant observer, that can in principle be compared to observations provided by
astronomers, in terms of three arbitrary metric functions. We also clarify the contribution of the
rotation and the dragging of inertial frames due to the gravitational field. We further point out
the difficulties we have when determining the metric functions from observations without making
reference to any theory of gravitational interactions, and comment on some possible ways to
overcome them.

\section{Stationary Axisymmetric Rotation curves}

We shall start by assuming that stars behave like test particles which follow time-like geodesics
of a rotating axially symmetric space-time associated with galactic discs which possess such a
symmetry. The most general line element for a space-time of this kind has the following form:

\begin{equation}
ds^2 = -e^{2\Phi} dt^2 + Q^2 dr^2 + R^2 \left[d\theta^2 + \sin^2\theta \left(d\varphi - Wdt\right)^2 \right],
\label{RGMS}
\end{equation}
where $\Phi, Q, R$ and $W$ are all functions of $r$
and $\theta$. We shall also consider two observers ${\cal O}_E$ and ${\cal O}_D$ with 4-velocities
$u^\mu_E$, $u^\mu_D$, respectively. Observer ${\cal O}_E$ corresponds to the light emitter (i.e.,
to the stars placed at a point $P_E$ of space-time), and ${\cal O}_D$ represents the detector at
point $P_D$, which is located far away from the light emitter and is ideally located at
$r\longrightarrow\infty$.

We further assume that stars move on the galactic plane and, thus, the polar angle can be fixed
$\theta=\pi/2$, so that $u^\mu_E = (U^t,U^r, 0, U^{\varphi})_E$, where $U^{\mu}=\dot x^{\mu}$ and
the dot stands for derivation with respect to the proper time of the particle (star).

On the other hand, the 4-velocity of the detector located ``far away" from the source, in a
stationary axisymmetric background, is given by $u^\mu_D = (U^t, U^r, 0, U^{\varphi})_D$ in the
language of the above mentioned coordinates and conventions. In this expression the $U^{\varphi}$
component of the 4-velocity $u^\mu_D$ accounts for the dragging of the observer at $P_D$ due to
the rotation of the galaxy. This effect must be taken into account when considering the
measurement of red/blue shifts of light signals emitted in our own galaxy or in galaxies ``close"
to ours, for instance. However, if the studied red/blue shifts correspond to galaxies located
``far away" from us, so that we can neglect the dragging effect, then we can consider that the
detector is static, i.e., that the ${\cal O}_D$'s 4-velocity is tangent to the static Killing
field ${{\partial} \over {\partial t}}$, and hence its 4-velocity is given by $u^\mu_D= (U^t, U^r,
0, 0)_D$. Later on we shall give a qualitative estimation of whether we can consider this ideal
limit (when the dragging effect can be neglected) in terms of the contribution of the angular
velocity of a given galaxy to the measured red/blue shift.

We further normalize the 4-velocity as usual $(u^\mu u_\mu=-1)$, a condition which renders the
following relation

\begin{equation}
 -1 = g_{tt} \left(U^t\right)^2 + g_{rr}\left(U^r\right)^2 +g_{\varphi\varphi}\left(U^\varphi\right)^2+ 2g_{t\varphi}U^tU^\varphi
 \label{mg}
\end{equation}
The stationary axisymmetric metric (\ref{RGMS}) possesses two commuting Killing vectors: the time-like
$\varepsilon^{\mu}=(1,0,0,0)$ and the rotational one $\psi^{\mu}=(0,0,0,1)$. The corresponding
conserved quantities are the energy and the angular momentum per unit of mass at rest of the test
particle and read

\begin{equation}
E = - g_{\mu\nu} \varepsilon^{\mu} u^{\nu} =-\left(g_{tt}U^{t}+g_{t\varphi}U^{\varphi}\right)
\label{E}
\end{equation}

\begin{equation}
L = g_{\mu\nu} \psi^{\mu} u^{\nu} = g_{\varphi\varphi}U^\varphi+g_{t\varphi}U^t
\label{ L}
\end{equation}
These relations are useful to obtain the expressions for $U^{t}$ and $U^{\varphi}$ in terms of the
metric and these conserved quantities:

\begin{equation}
U^t =
\frac{Eg_{\varphi\varphi}+Lg_{t\varphi}}{g_{t\varphi}^2-g_{tt}g_{\varphi\varphi}}=\frac{E-LW}{e^{2\Phi}}
\label{Ut}
\end{equation}

\begin{equation}
U^{\varphi}=-\frac{Eg_{t\varphi}+Lg_{tt}}{g_{t\varphi}^2-g_{tt}g_{\varphi\varphi}}=\frac{L}{R^2}+\frac{\left(E-LW\right)W}{e^{2\Phi}}
\label{L}
\end{equation}

By introducing these 4-velocity components (\ref{Ut}) and (\ref{L}) in the line element (\ref{mg}) we get

\begin{equation}
g_{rr}\left(U^{r}\right)^2=EU^t-LU^{\varphi} -1
\label{Ur2}
\end{equation}
or, after multiplying by $e^{2\Phi}$ we equivalently have

\begin{eqnarray}
 e^{2\Phi}Q^2\left(U^{r}\right)^2 + e^{2\Phi} + \frac{L^2e^{2\Phi}}{R^2} - \left(E-LW\right)^2   \nonumber \\
= e^{2\Phi}Q^2\left(U^{r}\right)^2 + V_{eff}(E,L,g_{\mu\nu})= 0
\label{Ur2axisim}
\end{eqnarray}

This equation resembles an energy conservation law for a non-relativistic particle with position
dependent mass moving in an effective potential that depends on $E$ and $L$. This equation can be
further reduced by considering that the orbits of stars are circular $U^r=0$:

\begin{equation}
V_{eff} = e^{2\Phi} + \frac{L^2 e^{2\Phi}}{R^2} - \left(E-LW\right)^2 = 0
\label{axisimcirc}
\end{equation}

On the other hand, the time-like geodesics that follow the stars orbiting around the galaxies can
be written as

\begin{equation}
\frac{d U^\mu}{d\tau}+\Gamma^\mu_{\rho\sigma}U^\rho U^\sigma = 0
\label{geodeqn}
\end{equation}
By computing the corresponding Christoffel symbols for the space-time metric
(\ref{RGMS}) and substituting their expressions into (\ref{geodeqn}) we obtain that geodesic
equations for $U^t$ and $U^\varphi$ are satisfied identically, while for the radial geodesic
equation we get:

\begin{eqnarray}
\frac{dU^r}{d\tau}+\frac{Q'}{Q}\left(U^r\right)^2+
\frac{1}{2Q^2}\left[\left(\frac{E\!-\!LW}{e^{2\Phi}}\right)^2\left(e^{2\Phi}\right)'\right. \nonumber \\
\left.-\frac{L^2}{R^4}\left(R^{2}\right)'+2L\left(\!\frac{E\!-\!LW}{e^{2\Phi}}\right)W'\right]=0,
\label{rgeodeqn1}
\end{eqnarray}
where $\tau$ is the proper time of the particle and $'$ denotes partial derivatives with respect
to $r$. By considering circular orbits we set again $U^r=0$ and obtain the following relation:

\begin{equation}
-\left(E-LW\right)^2\left(e^{-2\Phi}\right)'+\left(\frac{L^2}{R^2}\right)'+
2L\left(\!\frac{E-LW}{e^{2\Phi}}\right)W'=0 .
\label{circgeodescs}
\end{equation}

It turns out that this expression can be easily integrated yielding:

 \begin{equation}
 \frac{L^2}{R^2} - \frac{\left(E-LW\right)^2}{e^{2\Phi}} = k
\label{1int}
\end{equation}
where $k$ is an arbitrary constant. However, by comparing this result to the relation
(\ref{axisimcirc}), we see that it is in fact a fixed constant: $k=-1$. This situation also takes
place for the null geodesics of photons emitted by stars and detected by an observer located at
$r\longrightarrow\infty$: their geodesic equation is equivalent to the normalization condition of
their 4-momentum.

Once we have deleted the kinetic energy component in (\ref{Ur2axisim}) one gets the expression for
the effective potential on which the star moves (\ref{axisimcirc}). Thus, by deriving it with
respect to the radial coordinate and setting the result to zero, we basically impose the minimum
condition on the effective potential, a necessary condition to have circular orbits for the stars.
This condition leads to the following expression for $E$:

\begin{equation}
E =LW-\frac{\left(e^{2\Phi} + \frac{L^2 e^{2\Phi}}{R^2}\right)'}{2LW'}
\label{E1}
\end{equation}
whereas from equation (9) we get the following relation:
\begin{equation}
E = LW + e^{\Phi}\sqrt{1+\frac{L^2}{R^2}}
\label{E2}
\end{equation}
From both of these equations we see that the second term in the rhs of (14) must be positive definite,
leading to the following restriction:

\begin{equation}
\left(e^{2\Phi} + \frac{L^2 e^{2\Phi}}{R^2}\right)'  \grole 0, \qquad  W'  \leogr 0
\label{E-LW>0}
\end{equation}
where the last two inequalities must be satisfied simultaneously in the sense that the numerator
and the denominator of that term must have opposite sign for positive L.

Moreover, from (14) and (15) we obtain the expression for the angular momentum L:
\begin{equation}
L_{\pm}\!=\!\!
\sqrt{\!\frac{\left(W'\right)^2\!\pm\!\left|W'\right|\!\sqrt{\!\left(W'\right)^2\!+\!
\frac{2R'\left(e^{2\Phi}\right)'}{R^3}}\!-\!\frac{\left(e^{2\Phi}\right)'\left(\ln\!\frac{e^{\Phi}}{R}\!\right)'}{R^2}}
{\frac{2}{R^2}\left\{\left[\left(\frac{e^{\Phi}}{R}\right)'\right]^2 - \left(W'\right)^2 \right\}
}}, \label{Lpm}
\end{equation}
where the following restrictions must be fulfilled in order to have a real angular momentum:

\begin{equation}
(i) \quad \left(W'\right)^2 \ge - \frac{2R'\left(e^{2\Phi}\right)'}{R^3}, \label{restriction1}
\end{equation}
equivalently
\begin{equation}
\quad \left(W'\right)^2 \ge\left(e^{2\Phi}\right)'\left[\left(R\right)^{-2}\right]',
\label{restriction1b}
\end{equation}
this condition makes real the square root of the second term in the radicand's numerator;

\begin{eqnarray}
(ii) \qquad\left(W'\right)^2\pm\left|W'\right|\sqrt{\left(W'\right)^2+\frac{2R'\left(e^{2\Phi}\right)'}{R^3}}   \nonumber \\
 -\frac{\left(e^{2\Phi}\right)'\left(\ln\frac{e^{\Phi}}{R}\right)'}{R^2} \grole 0,
\label{restriction2a}
\end{eqnarray}

\begin{equation}
\left|\left(\frac{e^{\Phi}}{R}\right)'\right|  \grole \left|W'\right|,
\label{restriction2b}
\end{equation}
where, again, the last two inequalities must have the same sign simultaneously.

On the other side, for the energy $E$ we have from (\ref{E2}):

\begin{eqnarray}
\nonumber E_{\pm}\!=\!e^{\Phi}\!\!\sqrt{\!1\!+\!\frac{\left(W'\right)^2\!\pm\!
\left|W'\right|\!\sqrt{\!\left(W'\right)^2\!+\!\frac{2R'\left(e^{2\Phi}\right)\!'}{R^3}}\!-\!
\frac{\left(e^{2\Phi}\!\right)\!'\left(\ln\!\frac{e^{\Phi}}{R}\right)'}{R^2}}
{2\left\{\left[\left(\frac{e^{\Phi}}{R}\right)'\right]^2 -\left(W'\right)^2 \right\} }}  \nonumber \\
+W\!\!\sqrt{\!\frac{\left(W'\right)^2\!\pm\!\left|W'\right|\!\sqrt{\left(W'\right)^2\!+\!
\frac{2R'\left(e^{2\Phi}\right)\!'}{R^3}}\!-\!
\frac{\left(e^{2\Phi}\right)\!'\left(\ln\!\frac{e^{\Phi}}{R}\right)'}{R^2}}
{\frac{2}{R^2}\left\{\left[\left(\frac{e^{\Phi}}{R}\right)'\right]^2 - \left(W'\right)^2 \right\}
}}\!. \label{Energy}
\end{eqnarray}

Moreover, in order for the circular orbits of stars to be stable, we also need the second
derivative of the effective potential to be positive:

\begin{equation}
\left[e^{2\Phi}\left(1+\frac{L^2}{R^2}\right)\right]''+2L\left(E-LW\right)W''-2L^2\left(W'\right)^2 > 0,
\label{restricminptl}
\end{equation}
where $L$ and $E$ must be replace by their expressions (\ref{Lpm}) and (\ref{Energy}),
respectively. It is worth noticing that relations (\ref{Lpm}) and (\ref{Energy}), as well as the
condition (\ref{restricminptl}), can be reduced to the results obtained for the static spherically
symmetric approach reported in (\citealt{nucamendi}; \citealt{lake}) when $\theta=\pi/2$, $W=0$
and $R=r$ under a simple coordinate transformation. For instance, by taking $N=e^{\Phi}$ we have:

\begin{equation}
E^2=\frac{N^2}{1\!-\!r\partial_r N/N}\,, \quad L^2=\frac{r^3\partial_r N/N}{1\!-\!r\partial_rN/N}\,
\label{LakeEL}
\end{equation}
and
\begin{equation}
\quad  E=\frac{e^{\Phi}}{\sqrt{1-r\Phi'}}\,, \quad L=\frac{r\sqrt{r\Phi'}}{\sqrt{1-r\Phi'}}\,,
\label{LakeEL2}
\end{equation}
which are precisely the expressions obtained by (\citealt{nucamendi}) and (\citealt{lake}),
respectively, for the energy and angular momentum.

It is well known that rotation curves of spiral galaxies are inferred from the red and blue shifts
of the radiation emitted by stars that move in (nearly) circular orbits around the central region
of the galaxy  (\citealt{rubin1}; \citealt{rubin1a}; \citealt{rubin1b}) and that light signals
travel on null geodesics with tangent 4-momentum $k^{\mu}$. Here we shall make the assumption that
$k^{\mu}$ are restricted to lie in the galactic plane $\theta = \pi/2$, and we shall evaluate the
frequency shift of a light signal emitted from a star in a circular orbit represented by ${\cal
O}_e$ and detected by an observer represented by ${\cal O}_d$. Moreover, we shall suppose that the
galactic disc is edge-on directed towards the observer, a fact which implies that
$k^\theta=0$.\footnote{This fact is also taken into account by astronomers when reporting the
corresponding total blue or red shifts since they substract the contribution coming from the
inclination of the galactic disc from the plane $\theta = \pi/2$.}

The frequency shift associated to the emission and detection of light signals is given by

\begin{equation}
1+z = \frac{\omega_e}{\omega_d} \,\,\,\,,
\label{z}
\end{equation}
where the frequency of a photon
measured by an observer with proper velocity $u^\mu_C|_{P_C}$ reads
\begin{equation}
\omega_C=-k_\mu u^\mu_C|_{P_C}\,\,\,\,,
\label{freq}
\end{equation}
and the index $C$ refers to the emission (e) or detection (d) at the corresponding space-time
point $P_C$. When the observer is comoving with the particle we have
\begin{equation}
\omega_e = -k_\mu u^\mu_e\,\,\,\,.
\label{freqcomovobs}
\end{equation}
Thus, a photon emitted at point $P_C$ in the galactic plane possesses a 4-momentum
$k_C^\mu=\left(k^t,k^r,0,k^\varphi\right)_C$. The corresponding conserved quantities along the
null geodesics of light signals read

\begin{equation}
E_\gamma = - g_{\mu\nu} \varepsilon^{\mu} k^{\nu} =-\left(g_{tt}k^{t}+g_{t\varphi}k^{\varphi}\right) \,\, ,
\label{Egamma}
\end{equation}

\begin{equation}
L_\gamma = g_{\mu\nu} \psi^{\mu} k^{\nu} = g_{\varphi\varphi}k^\varphi+g_{t\varphi}k^t \,\, ,
\label{Lgamma}
\end{equation}
whereas the normalization condition for the 4-momentum $k^\mu k_\mu=0$
is

\begin{equation}
0 = g_{tt} \left(k^t\right)^2 + g_{rr}\left(k^r\right)^2 +g_{\varphi\varphi}\left(k^\varphi\right)^2+ 2g_{t\varphi}k^tk^\varphi \,\, .
\label{k2=0}
\end{equation}
and leads to the following expression for the $k^r$ in terms of the metric, the conserved
quantities $L_\gamma$, $E_\gamma$, and $k^t$ and $k^\varphi$:

\begin{equation}
g_{rr}\left(k^r\right)^2 = E_\gamma k^t - L_\gamma k^\varphi = \frac{g_{\varphi\varphi}E_\gamma^2 + 2g_{t\varphi}E_\gamma L_\gamma +g_{tt}L_\gamma^2}{g_{t\varphi}^2-g_{tt}g_{\varphi\varphi}}.
\label{kr2}
\end{equation}

There are two frequency shifts which correspond to the maximum and minimum values of $\omega_E$
associated with light propagation in the same and the opposite direction of the motion of the
signal emitter, in other words, the frequency shifts of a receding and an approaching star,
respectively. As we shall see later, these max/min values of the frequency shifts are reached for
stars whose position vector ${\bf r}$, with respect to the galactic center, is perpendicular to
the detector's line of sight, i.e., along the plane where $k^r=0$ for the observer \citep{lake}.

From the constant character along the geodesics of the product of the Killing vector field
$\varepsilon^{\mu}$ with a geodesic tangent (\ref{Egamma}), together with the frequency definition
(\ref{freq}) and taking into account that at ($r\longrightarrow\infty$), the 4-velocity of the
detector is given by $u^{\mu}_d=(U^t,0,0,U^\varphi)_d$, we get the following expressions for the
detector's frequency

\begin{eqnarray}
   \omega_d &=& -k_\mu
u^\mu_d \nonumber \\
&=&-\left.\left(g_{tt}U^tk^t+g_{t\varphi}U^tk^\varphi +g_{\varphi t}U^\varphi
k^t+g_{\varphi\varphi}U^\varphi k^\varphi\right)\right|_d \nonumber \\
  &=& \left.\left(E_\gamma U^t - L_\gamma
U^\varphi\right)\right|_d\,\,,
\label{freqdetec}
\end{eqnarray}
and the emitter's frequency

\begin{eqnarray}
 \omega_e &=& \left.-k_\mu
u^\mu\right|_e \nonumber \\  &=&-\left(g_{tt}U^tk^t+g_{t\varphi}U^tk^\varphi+g_{\varphi
t}U^\varphi k^t+g_{\varphi\varphi}U^\varphi k^\varphi \right. \nonumber \\
&+&\left.\left. g_{rr}U^rk^r\right)\right|_e = \left.\left(E_\gamma U^t-L_\gamma
U^\varphi-g_{rr}U^rk^r\right)\right|_e\, \label{freqem}
\end{eqnarray}

Therefore, with the aid of the frequency shift (\ref{z}) we have for arbitrary star orbits

\begin{equation}
1+z = \frac{\omega_e}{\omega_d} = \frac{\left.\left(E_\gamma U^t-L_\gamma
U^\varphi-g_{rr}U^rk^r\right)\right|_e}{\left.\left(E_\gamma U^t - L_\gamma
U^\varphi\right)\right|_d} \,\,\,\,.
\label{zgamma}
\end{equation}
Since the star orbits we are considering are circular, then

\begin{equation}
 1+z =
\frac{\omega_e}{\omega_d} = \frac{\left.\left(E_\gamma U^t-L_\gamma
U^\varphi\right)\right|_e}{\left.\left(E_\gamma U^t - L_\gamma U^\varphi\right)\right|_d} =
\frac{U^t_e - bU^\varphi_e}{U^t_d - b U^\varphi_d}\,\,\,\,,
\label{zcircorbits}
\end{equation}
where we have introduced $b\equiv\frac{L_\gamma}{E_\gamma}$ -- the impact parameter at infinity,
and, hence, $|b|$ represents the radial distance at any side of the observed center of the galaxy,
where $b=0$. Since we shall consider red/blue shifts either side of the central value $b=0$, it is
convenient to express (\ref{zcircorbits}) as
\begin{equation}
1+z_{\epsilon} = \frac{\left.\left(U^t-\epsilon |b|
U^\varphi\right)\right|_e}{\left.\left(U^t-\epsilon |b| U^\varphi\right)\right|_d}
\,\,\,\,,
\label{zmodulb}
\end{equation}
where $\epsilon=\pm 1,$ and to compute the following quantity:

\begin{equation}
1+z_c = \frac{U^t_e}{U^t_d} =
\frac{\left.\left(\frac{E-LW}{e^{2\Phi}}\right)\right|_e}{\left.\left(\frac{E-LW}{e^{2\Phi}}\right)\right|_d}\,\,,
\label{zatbnull}
\end{equation}
where $L$ and $E$ are given by (\ref{Lpm}) and (\ref{Energy}), respectively. This is the
gravitational red shift of the center of the galaxy measured by an observer located at
$r\longrightarrow\infty$. Since in order to consider red/blue shifts either side of the central
value $b=0$, we need to subtract this quantity from (\ref{zmodulb}), we define red and blue shifts
as follows
\begin{equation}
 Z_{red}\equiv z_+-z_c = \frac{\left(U^t_e U^\varphi_d-U^t_d
U^\varphi_e\right)}{U^t_d\left(U^t_d-|b|U^\varphi_d\right)}|b|\,\,\,\,,
\label{Zr}
\end{equation}

\begin{equation}
 Z_{blue}\equiv z_c-z_- = \frac{\left(U^t_e U^\varphi_d-U^t_d
U^\varphi_e\right)}{U^t_d\left(U^t_d+|b|U^\varphi_d\right)}|b|\,\,\,\,,
\label{Zb}
\end{equation}
corresponding to receding and approaching stars, respectively. It is obvious that now $Z_{red}\ne
Z_{blue}$ (since $z_+-z_c \ne z_c-z_-$). These quantities are the general red and blue shifts,
respectively, experimented by light signals traveling along null geodesics and emitted by
circularly orbiting stars around the center of a galaxy to a distant observer. It is worth
mentioning that in this formula we have dropped the gravitational red shift of the center of the
galaxy, a fact that it is indeed taken into account by astronomers when reporting their observed
data.

In the special case in which the detector is located ``far away" enough from the source of
information, i.e., if the contribution of the dragging of the detector's inertial frame due to the
rotation of the system, encoded in $U_d^\varphi$, is negligible in comparison to the contribution
coming from the $U_d^t$ component, in other words, if $U_d^\varphi<<U_d^t$, then the detector can
be considered static at $r\longrightarrow\infty$ and its 4-velocity will be given by
$u^\mu_D=e^{2\Phi(\infty)}\delta^\mu_t$. This fact can be understood as well from the following
analysis: since we are considering the limit in which $U_d^\varphi<<U_d^t$, and taking into
account that $u^\mu_d=\left.\frac{d x^\mu}{ds}\right|_d$ then

\begin{equation}
\frac{U_d^\varphi}{U_d^t}=\frac{d\varphi}{dt}\equiv \Omega_d << 1,
\label{OmegaD}
\end{equation}
 where $s$ is the proper time of the orbiting particle (star) and $\Omega_d$ is the angular velocity measured
by the detector at $r\longrightarrow\infty$.

Thus, in the special case in which $U_d^\varphi<<U_d^t$ (or $\Omega_d << 1,$) the general red/blue
shifts (\ref{Zr}) and (\ref{Zb}) reduce to the effective (quasi-static) red/blue shift defined by
the following expression:

\begin{equation}
 Z = -\frac{U^\varphi_e}{U^t_d}|b| =
-\frac{\left.\left[\frac{L}{R^2}+\frac{(E-LW)W}{e^{2\Phi}}\right]\right|_e}
{\left.\left(\frac{E-LW}{e^{2\Phi}}\right)\right|_d}|b|\,,
\label{Zqstatic}
\end{equation}
which is symmetric with respect to the center of the galaxy as in (\citealt{nucamendi}; \citealt{lake}). It is
worth noticing that even when the dragging of the detector's inertial frame is neglected, the
differential rotation encoded in $W$, still plays a nontrivial role in the relation
(\ref{OmegaD}). Hence we have
\begin{equation}
Z^2 =
\frac{\left.\left[\frac{L}{R^2}+\frac{(E-LW)W}{e^{2\Phi}}\right]^2\right|_e}
{\left.\left(\frac{E-LW}{e^{2\Phi}}\right)^2\right|_d}b^2\,.
\label{Zqstatic2}
\end{equation}

We further need to take into account the light bending due to the gravitational field generated by
the rotating galaxy, in other words, we need to construct a mapping between the impact parameter
$b$ and the location of the star $r$ given by its vector position ${\bf r}$ with respect to the
center of the galaxy, i.e., the mapping $b(r)$. Following (\citealt{nucamendi}; \citealt{lake}), we shall choose
the maximum value of $Z^2$ at a fixed distance from the observed center of the galaxy (at a fixed
$b$). From (\ref{Zqstatic2}) it follows that if the function factor that multiplies $b^2$ is
monotone decreasing with increasing $r$, then the maximum observed value of $Z^2$ corresponds to
the minimum value of $r$ along the null geodesic of the photons. This minimum value of $r$
corresponds to the position of the star either side of the center of the galaxy, lying on the
plane perpendicular to the detector's line of sight, i.e., on the plane where $k^r=0$ for the
observer located at $r\longrightarrow\infty$. Thus, from (\ref{Zqstatic2}) it follows that the
squared impact parameter $b^2$ must also be maximized; this quantity can be calculated from the
geodesic equation of the photons (or, equivalently, from the $k^\mu k_\mu =0$ relation taking into
account that $k^r=0$) and is given by

\begin{eqnarray}
 b_\pm =
\frac{-g_{t\varphi}\pm\sqrt{g_{t\varphi}^2-g_{tt}g_{\varphi\varphi}}}{\left(g_{tt}\right)} =
-\frac{R^2 W\pm R e^{\Phi}}{e^{2\Phi}},   \nonumber \\
  \qquad b_\pm^2 = \frac{\left(R^2 W\pm R
e^{\Phi}\right)^2}{e^{4\Phi}}.
\label{b}
\end{eqnarray}

Since we look for the maximum value of $b$, then we should choose either $b_+$ or $b_-$, depending
on the sign and magnitude of the product $RW$ with respect to $e^{\Phi}$, in such a way that $b$
is maximum. Finally, the monotone decreasing condition of the factor that multiplies $b^2$ in
(\ref{Zqstatic2}) imposes the following restriction

\begin{equation}
 \left[\ln\left(\frac{L}{R^2}+\frac{(E-LW)W}{e^{2\Phi}}\right)_e\right]'<
\left[\ln\left(\frac{E-LW}{e^{2\Phi}}\right)_d\right]'\,.
\label{mondeccond}
\end{equation}
Thus, the mapping $b(r)$, responsible for the gravitational light bending, is given by (\ref{b})
under the condition (\ref{mondeccond}).

We should also mention here that the relations (\ref{b}) and (\ref{mondeccond}) reduce to the
results obtained by (\citealt{nucamendi}; \citealt{lake}) when $W=0$ and $R=r$:

\begin{equation}
 b^2=\frac{r^2}{e^{2\Phi}}\qquad \mbox{and} \qquad
\left(\ln\frac{L}{r^2}\right)'< \left(\ln\frac{E}{e^{2\Phi}}\right)'\,,
\label{b2+mondeccond}
\end{equation}
respectively.

As mentioned above, observations are reported as

\begin{equation}
Z=v(b)-v(b=0)
\label{eq1.45}
\end{equation}
where $Z$ is given by (\ref{Zqstatic2}), supplemented by the light bending mapping (\ref{b}).

If one succeeds in determining the three unknown metric functions $\Phi$, $R$ and $W$ from
observational data, this would imply that the dynamics of light signals is determined by the
geodesics of a stationary axisymmetric metric, independently of the assumption of the dynamics of
the geometry (of a theory of gravitational interactions) or of the nature of dark matter, in the
case that the latter is needed. However, the task of solving for $\Phi$, $R$ and $W$ is not
trivial at all compared to the spherically symmetric case considered in (\citealt{nucamendi}; \citealt{lake})
where there was just one unknown metric function to be determined.

A way of determining the above mentioned three arbitrary metric functions from observations
consists in making use of the empiric Persic and Salucci's universal formula for rotation
curves in the halo region of a galaxy \citep{persic}, see also \citep{salucci2}:
\begin{equation}
 Z_i= \frac{\alpha_i
b^2}{b^2+\beta_i},
\label{zi}
\end{equation}
where the constants $\alpha_i$ and $\beta_i$ correspond to the description of the central, red and
blue shifts defined in (\ref{zatbnull})--(\ref{Zb}), and are determined from observations of a
given galaxy. An alternative way to proceed is to consider just the red and blue shifts,
(\ref{Zr}) and (\ref{Zb}), related to this formula and make use of the relation (\ref{E1}) (or
(\ref{E2}) if possible) in order to complete the system of equations for $\Phi$, $R$ and $W$.
Thus, in principle, we can conform a system of three nonlinear differential equations of first
order for three unknown metric functions. From the expressions found for the red/blue shifts it is
obvious that the difficulties that arise in solving such a system are not trivial at all.

There is another kind of universal rotation curve formula for flat spiral galaxies that can be adjusted
to observed data  (\citealt{GPPRC}) and can be used to determine $\Phi$, $R$ and $W$ upon setting it to
the red/blue shift language. Moreover, both of these universal rotation curve formulas could be
used in a combined way in order to determine the metric functions $\Phi$, $R$ and $W$ from
observations for one and the same sample of galaxies.

One more way to find these functions consists of approaching the galactic rotation curves problem
in the post-Newtonian approximation in the spirit of \citep{RCAP}, where the authors constructed
analytical models that allow one to compute the first general relativistic corrections to the
matter distributions and gravitational potentials for stationary systems with axial symmetry. It
is worth mentioning that the main modifications in their approach appear far from the galaxy
cores, a result that seems to be consistent with the predictions of (\citealt{CT}; \citealt{CTb};
\citealt{CTc}; \citealt{CTd}; \citealt{CTf}).

 We would like to point out that one can make a combined use of the red/blue shifts of galactic
rotation curves and gravitational lensing as proposed in (\citealt{kar};\citealt{visser}). However, as stated in
\citep{visser}, till now the combined measured data come from different distance scales (red
shifts): most high quality rotation curves are available for galaxies with a low to intermediate
red shift of up to $z\sim 0.4$, while gravitational lenses are easier to detect at intermediate to
high red shifts $z\ga 0.4$. Therefore, both kinds of data are available for the same galaxy,
but at different radii and, hence, they are not comparable. Thus, it is still difficult to take
advantage from both sets of observations simultaneously. This situation will improve in the future
when observations with a higher resolution will be carried out. In summary, more precise
observational data are needed in order to benefit from the stationary axisymmetric approach to the
galactic rotation curves problem within this context.

Finally, the stationary axisymmetric formalism presented here could be applied to a wider range of
phenomena 
like binary systems, accretion discs of rotating black holes and
active galactic nuclei where the size of the effects would be less restrictive.

\bigskip

\noindent{\bf Acknowledgments}

\noindent All authors express their gratitude to the people of Mexico for their kind support of
fundamental research. OC is grateful to INFN Bologna for hospitality and support while parts of
this work were completed. AHA and UN thank Pedro Col\'\i n and Daniel Sudarski for fruitful and
illuminating discussions and the Physics Department of CEFyMAP, UNACH for hospitality. AHA is also
grateful to the staff of the ICF, UNAM for hospitality. The work of OC was partly funded by
SEP--PROMEP/103.5/11/6653. AHA acknowledges support from the PAPIIT-UNAM grant No. IN103413-3 
{\it Teor\'ias de Kaluza-Klein, inflaci\'on y perturbaciones gravitacionales.} AHA and ES have 
benefited from a CONACYT grant No. I010/393/2011 C.525/2011. UN acknowledges support from the 
CIC--UMSNH Project No. 4.8. All authors thank SNI and PROMEP as well.

\label{lastpage}

\end{document}